Title: Better artificial intelligence does not mean better models of biology

Authors: Drew Linsley[1], Pinyuan Feng[2], and Thomas Serre[1]
Affiliation: [1]Brown University, [2]Columbia University

**Abstract**

Deep neural networks (DNNs) once showed increasing alignment with primate perception and neural responses as they improved on vision benchmarks, raising hopes that advances in AI would yield better models of biological vision. However, we show across three benchmarks that this alignment is now plateauing—and in some cases worsening—as DNNs scale to human or superhuman accuracy. This divergence may reflect the adoption of visual strategies that differ from those used by primates. These findings challenge the view that progress in artificial intelligence will naturally translate to neuroscience. We argue that vision science must chart its own course, developing algorithms grounded in biological visual systems rather than optimizing for benchmarks based on internet-scale datasets.

**Highlights**

- In the past, deep neural networks (DNNs) showed increasing alignment with primate neural responses as their object recognition accuracy improved.
- Across three different neural and behavioral datasets, we find that state-of-the-art DNNs with human-level accuracy are now worsening as models of primate vision.
- Today's DNNs that are scaled-up and optimized for artificial intelligence benchmarks achieve human (or superhuman) accuracy, but do so by relying on different visual strategies and features than humans.
- DNNs can still be designed to learn more human-like strategies and behavior, but continued improvements will not come for free from internet data.
- Vision science must break from artificial intelligence and develop new deep learning approaches that are tailored to biological vision.

**The transformative promise of deep learning for vision science.**

One year after AlexNet—the **deep neural network** (DNN) that sparked the modern artificial intelligence revolution—Yamins and colleagues [1,2] published a pivotal finding: units in DNNs honed for perceptual tasks like object recognition responded to images in similar ways to real neurons in primate inferotemporal (IT) cortex. This discovery transformed vision science. Suddenly, DNNs weren't just toy models for categorizing dogs and cats, they were now models of biology that could settle long-standing debates about neural computations and their links to perception. For example, the observation that DNNs pre-trained for object classification better predicted IT responses provided computational backing to the theory of core object recognition—the idea that primate visual circuitry evolved for rapid and invariant object recognition [3]. Perhaps most importantly, these newfound capabilities of DNNs implied that optimizing DNNs for computer vision and artificial intelligence tasks might unlock the secrets of biological vision in the process. Advances in artificial intelligence through engineering could lead the way, and biological discovery would follow as a byproduct.

Over the decade since AlexNet, DNNs have advanced dramatically and now rival or surpass humans on benchmarks spanning vision, language, reasoning, and planning [4–7]. This progress has almost entirely been driven by an exponential scale-up of DNNs, a trend that was catalyzed by the introduction of **Transformers** [8]—attentional circuits that are designed for the parallel processing capabilities of GPUs—and the observation that DNN performance on nearly all benchmarks improves predictably as the number of model parameters and amount of data used for training increases [4–7,9,10]. Nevertheless, today's state-of-the-art DNNs including frontier models like OpenAI's GPT-4o, Anthropic's Claude 3, and Google Gemini 2—systems estimated to contain billions of parameters and trained on large proportions of the internet—still behave in strange ways; for example, stumbling on problems that seem trivial to humans while excelling at complex ones [11–16]. This chasm between the human-level accuracy but alien behaviors of today's DNNs raises a critical question for vision science: has the relentless engineering of DNNs to improve on artificial intelligence benchmarks continued to advance our understanding of brains and behavior, or has it steered these systems away from biological principles altogether?

**How task-optimized deep learning reshaped vision science.**

The task-optimized deep learning [17,18] method pioneered by Yamins and colleagues is now a standard tool for modeling primate vision. This approach is powerful for multiple reasons. It is the state-of-the-art modeling technique for system identification and predicting primate neural responses to images, which could be useful for device development [19–22] and *in silico* studies of function [23–25]. Perhaps more importantly, task optimization can be used to glean insights into the principles that shape visual circuitry. For example, Yamins and colleagues found that activities from a DNN trained solely for object classification correlate more strongly with neural responses than one trained end-to-end for neural prediction [1,2]. This result implied that object classification was a guiding principle for the organization of the visual system, especially circuits in **V4** and **Inferotemporal Cortex (ITC)** [26–41]. Others have found that the success of other pre-training objectives, such as those for **self-supervision** [28,29,42–44], makes the story less clear cut.

Task-optimization, and specifically training DNNs for image-based tasks like object classification, has also proven to be an effective strategy for modeling human perception. Most prominently, it has repeatedly been shown that the alignment between human and DNN object recognition decisions in psychophysics experiments increases with model accuracy on computer vision benchmarks [45–49]. This convergence is more than just a consequence of fewer mistakes; even the patterns of errors produced by these DNNs have grown to resemble humans [48,50]. Object classification-optimized DNNs are capable of predicting human judgements across diverse tasks, including tests of local vs. global processing biases [51], the perceived semantic similarity of object images [52,53], bottom-up saliency of images [54,55], and the 3D properties of objects and scenes [11,56,57]. However, even with task optimization, there are still multiple features of human perception that are challenging to model without additional tricks, such as perceptual phenomena like illusions [58–64] and metamers [65–68].

Despite the success of task-optimized deep learning for primate visual systems, task pre-training has not been necessary for modeling simpler neural systems. For example, salamander retina circuits can be approximated by DNNs trained end-to-end "from scratch" to predict electrophysiological recordings of retinal ganglion cells (RGC). Probing the trained DNNs' units subsequently provided mechanistic explanations of poorly understood RGC phenomena like motion anticipation and reversal [69–72]. Similarly, end-to-end training has been successful for predicting calcium imaging recordings of neurons in mouse early visual cortex [27,31,73–76]. The resulting models have also made precise control over neural function possible and enabled new types of neuroscience studies; for example, bringing unprecedented clarity into the circuits underlying contextual effects in neural responses [74,75].

Inspired by the success of engineering benchmarks at advancing artificial intelligence, the visual neuroscience community has developed its own benchmarks to systematically evaluate the prediction accuracy of models of biological vision [77,78]. A particularly influential initiative is Brain-Score, an online benchmark with a living leaderboard that allows anyone to submit and evaluate models on private data from dozens of neural and psychophysics experiments [79,80]. By tracking how DNNs perform on biological benchmarks as they improve on engineering ones, Brain-Score offers essential insights into how advances in artificial intelligence have translated to biological vision.

**What forms of task optimization lead to better models of biological vision in today's DNNs?**

Optimizing DNNs on image tasks like object classification or self-supervision has become an increasingly unreliable strategy for modeling biological vision. We saw the first hint of this decline in the **Brain-Score** benchmark [33,81], which features the same experimental dataset [82] as Yamins and colleagues, but is updated to include newer and more advanced DNNs. For each of these models, Brain-Score reports their **ImageNet** [83] accuracy and the peak alignment between their units and primate ITC neurons (using dimensionality reduction then linear regression [79,80], Fig. 1A). Over recent years, a striking trend has emerged: DNNs continued to improve as models of ITC until they reached around 70% accuracy on ImageNet, at which point they explained approximately half of the neural variance. Beyond this point, their **neural alignment** scores have declined even as their ImageNet accuracies have improved. In other words, image task-optimization of DNNs for object classification may no longer be an effective strategy for building better models of vision.

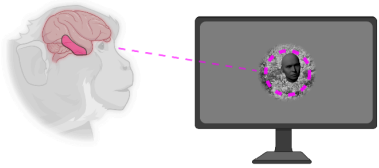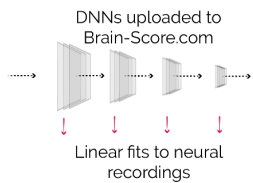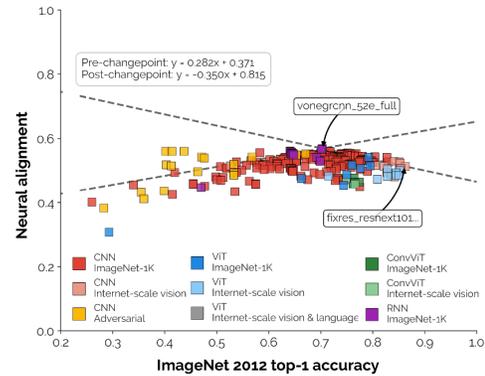
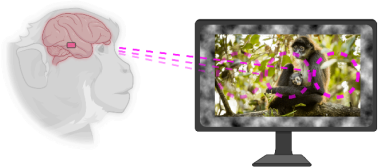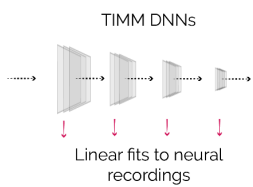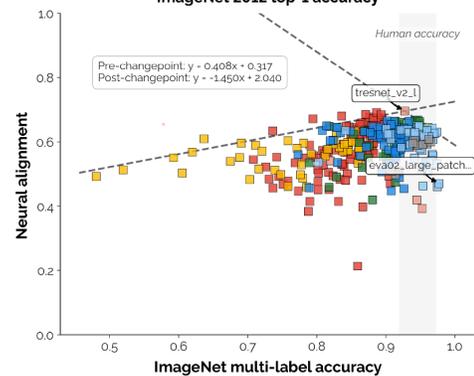
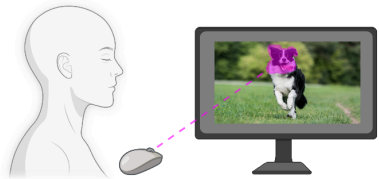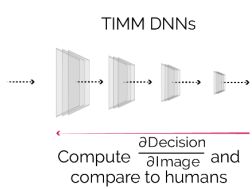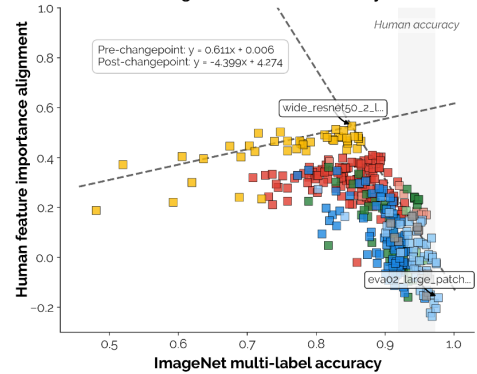

**Figure 1. DNNs have grown worse as models of biological vision as they have improved at object recognition. (A)** DNNs optimized for image tasks like object recognition are the standard for predicting image-evoked responses of neurons in the primate inferotemporal cortex (ITC) [82]. Units in each layer of an image task-optimized DNN are linearly fit to a training set of recordings, and then the best-fitting layer is used to make predictions on independent recordings [79]. While there once was a strong and linear relationship between the object recognition accuracy of a model and its ability to predict ITC responses to images, the sign of that correlation has flipped as models have surpassed 70% accuracy on ImageNet. Each dot represents a different model. Dashed lines denote trends of models on the pareto frontier of neural alignment (i.e., best-case biological predictions). The model representing the changepoint for the trends is the VOneGRCNN-52e-full [84], a recurrent convolutional neural network. **(B)** A similar pattern of results was found in another study of ITC [33,85], which sought to map the firing rates of neurons to different parts of the same image, yielding neural activity "heatmaps" for images. The initially positive relationship between DNN object classification accuracy and neural predictivity flipped with the release of the TResNet [86]. The grey patch denotes human accuracy on this version of ImageNet [87]. **(C)** One explanation for this flipped trend is that DNNs have learned to base their decisions on visual features that are different from those of humans [88–91]. In this study, visual feature importance maps were derived from human participants who identified parts of images they considered diagnostic for recognition. DNN feature importance maps were then derived using standard XAI methods based on the model's decision gradient. The alignment between humans and DNNs was measured by computing the correlation between these maps. The changepoint is an Adversarially-trained Wide ResNet50 [92].

To dissect this trend of image-classification optimized deep learning and understand how widespread it is, we turned to a new set of experimental recordings, in which monkeys fixated on different locations in natural images as neurons in their posterior ITC were being recorded ([33], Fig. 1B). We then benchmarked the neural alignment of a large and diverse set of DNNs from the Pytorch Image Model library (TIMM [93]). The TIMM library contains a decade's worth of progress in computer vision, spanning the architectures (**Convolutional Neural Networks**/CNN, **Vision Transformers**/ViT, and hybrids of these two) and pre-training approaches (e.g., first training on ImageNet-21K [94], using self-supervised approaches like DINO[95], or using image-language modeling approaches like CLIP [96]) that have represented the state-of-the-art over that period. Importantly, every model is implemented using a standardized approach and fine-tuned on ImageNet [83], which contrasts with Brain-Score, where models of unknown provenance are submitted pseudonymously.

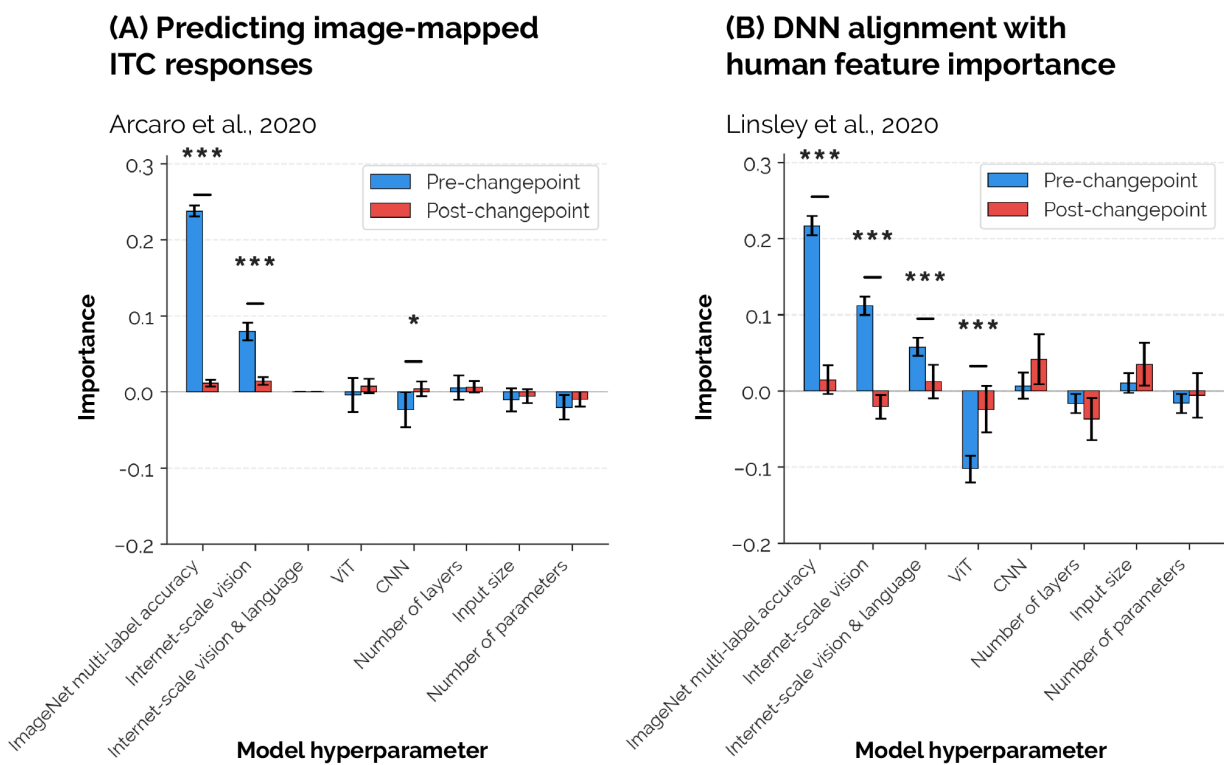

**Figure 2. As strategies for developing DNNs have changed over the past decade, so too have the model factors that drive their predictions of biological vision.** ANOVAs were used to estimate the relative importance of DNN hyperparameters for predicting image-mapped ITC responses or human feature importance in images: for example, if the model was pre-trained on Internet-scale data, and if its architecture had transformer (ViT) layers, convolutional layers (CNN), etc. Each model comes from the TIMM library and was ultimately trained (or fine-tuned) for ImageNet classification. Significance tests denote differences in the importance of a parameter for explaining performance in the pre-changepoint group vs. the post-changepoint group. The changepoint model for "Predicting image-mapped ITC responses" was the TResNet [97,98], and for "DNN alignment with human feature importance" was an Adversarially-trained Wide ResNet 50 [92]. Error bars denote S.E.M. *: $p < 0.05$, ***: $p < 0.001$.

We evaluated hundreds of DNNs from TIMM on two different objectives. First, we measured accuracy on a multilabel version of ImageNet that was benchmarked against humans in a year-long experiment [87]. In addition to the extraordinary effort it took to measure the accuracy of five human participants on 80% of the ImageNet validation set, this multilabel version also resolved a key issue in the original ImageNet, in which the multiple objects present in some images had confounded earlier attempts to evaluate standard single-object classification accuracy. Second, we measured the models' ability to predict primate ITC responses [82] using the Brain-Score procedure [79,80]. ImageNet performance of a large proportion of DNNs in this set rivaled or surpassed humans: 30% (111) had human-level object classification accuracy (scoring within the human confidence interval) and 1% (4) outperformed humans (the EVA class of models [99]; Fig. 1B). At the same time, these state-of-the-art DNNs had also devolved into poor models of primate vision. The superhuman EVA-02 [99], released in 2023, was as aligned with ITC as the MobileNetV3 [100] from 2019. The TResNet—a GPU performance-optimized implementation of the original ResNet [97,98]—represented an inflection point for DNNs where the previously positive correlation between task performance and neural alignment turned negative. Follow-up analyses comparing model performance on other benchmarks, like the original ImageNet or more challenging versions [101–103], showed similar reversals. Other controlled analyses have come to similar conclusions [80].

Using TIMM models in this analysis allowed us to analyze the relationship between the design of DNNs and their declining neural alignment. We split DNNs into pre- and post-TResNet (lower vs. higher accuracy), then used ANOVAs to estimate how each group's architectural choices, training recipes, and ImageNet accuracy affected neural alignment. Architectural choices were whether the model was a ViT, a CNN, the number of layers, the size of its input, and the number of parameters. Training recipes were if the model was trained just on ImageNet, or also pretrained on Internet-scale image data (e.g., ImageNet-21K) or Internet-scale vision and language data (e.g., LAION-5B [104]). The neural alignment of the lower-accuracy group was significantly correlated with ImageNet performance and training dataset size, echoing earlier findings [1,2,79,80] (Fig. 2A). However, there were no model design choices for the higher-accuracy group that explained neural alignment differences. This includes the number of parameters or layers in a DNN, the diversity of its training data, or its architectural family. These results suggest that the very computational strategies that have propelled DNNs to superhuman accuracy are uncorrelated with those that govern primate vision.

**Why does task optimization now lead to worse models of biological vision?**

One possible explanation for the declining effectiveness of task-optimized deep learning is that DNNs have begun to discover visual strategies that are challenging for biological visual systems to exploit as they have been scaled up. Early hints of this problem came from Ullman and colleagues [105], who found that early DNNs, like AlexNet and VGG [106], came to their object recognition decisions differently than humans. Humans exhibited an all-or-nothing reliance on diagnostic features, and the excision of a small number of pixels in an image was enough to make an image no longer recognizable to a human observer. DNNs did not show the same tendency. These results were surprising and exciting, demonstrating a new dimension for comparing human and DNN vision. At the same time, the study used a very small dataset and was limited in multiple ways [107], leaving open questions of whether or not these differences between humans and DNNs were mere experimental quirks or more fundamental differences.

To answer this question, we developed ClickMe, a scalable, gamified platform designed to collect high-resolution feature importance maps for large natural image datasets [88,89,108]. In ClickMe, participants teach virtual students (DNNs) to recognize objects by highlighting diagnostic regions (Fig. 1C). The resulting maps highlighted visual features people thought were important, were highly consistent across individuals, and were significantly more diagnostic for human object classification than bottom-up saliency [88]. We collected these maps for a large proportion of ImageNet, then compared their alignment with **attribution maps** from DNNs [109–112], computed as smoothed gradients of model decisions with respect to input pixels, by correlating human and DNN maps for each image.

This experiment revealed a growing gap between artificial and human vision: as DNNs have advanced beyond human accuracy, the features they have learned to use for recognition have progressively grown more different than those used by humans. For example, an adversarially-trained ResNet50 achieved the highest alignment with humans, capturing ~50% of the variance in human feature importance maps. The alignment of models with greater accuracy than this ResNet dropped precipitously as they began to exploit different features (e.g., background textures, global statistics, even labels; see Fig. 3 for examples) than humans [33,89,113]. ANOVAs confirmed these trends and showed that significant correlations between the data diets, training routines, and human-alignment of less-accurate DNNs (prior to the adversarially-trained wide ResNet highlighted in Fig. 1C) have disappeared in more accurate models (Fig. 2B; following the same procedure outlined above for Fig. 2A). These results underscore the fundamental shift in DNN optimization strategies that have taken place over recent years: the scale-up of DNNs to human-level performance has come at the cost of modeling biology. Today's state-of-the-art DNNs do not achieve their performance by discovering mechanisms honed by evolution to support intelligent behavior. Instead, they discover entirely novel—and often inscrutable—strategies which may be highly valuable for performance in non-ethological visual recognition tasks such as biomedical image analysis, but less so for understanding brains and behavior.

**Task-optimized DNNs may no longer produce better biological models. What will?**

The magnificent promise of task-optimized deep learning was in its suggestion that all we needed was more data, compute, and engineering to reverse-engineer vision. The now waning effectiveness of this approach represents an opportunity for vision scientists to rethink how we model brains. The first question we must ask is whether DNNs are the right modeling framework to begin with [65].

We sought to answer this question by investigating if we could force DNNs into alignment with primate vision by training them to use similar visual features as humans during object classification (Fig. 1C). Our Harmonization method guides DNNs to have decision attribution maps that are as close as possible to human feature importance maps (e.g., collected with ClickMe) during ImageNet training [114]. Harmonizing DNNs aligned their visual strategies and object classification decisions with humans: it led to more human-like sensitivity to adversarial attacks [113] and significantly improved neural alignment with ITC [33] (Fig. 3). These experiments indicate that DNNs are still a viable approach for building better models of biological vision. The challenge ahead isn't simply to achieve better alignment with biological vision through engineering with new and bigger datasets like ClickMe, but to discover the fundamental principles that shape primate visual systems—principles that could guide the development of DNNs that perceive the world like humans from the outset.

## ClickMe

Linsley et al., 2020

## ViT Small

Dosovitskiy et al., 2021

## Harmonized ViT Small

Fel* et al., 2022

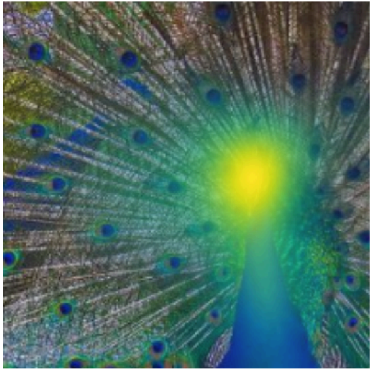
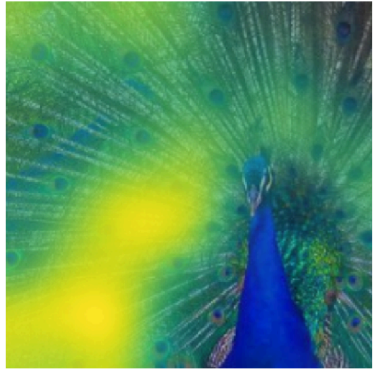
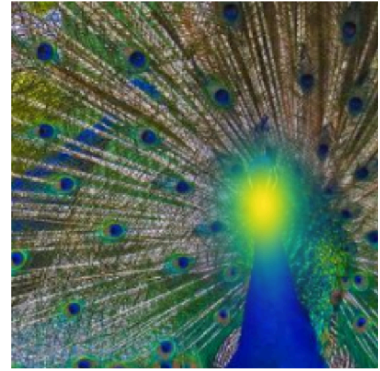
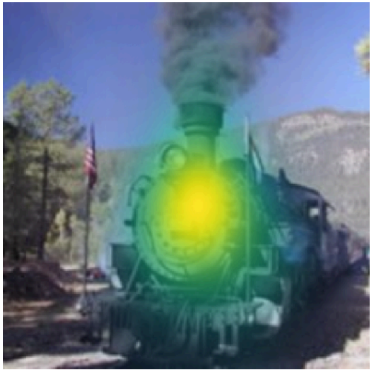
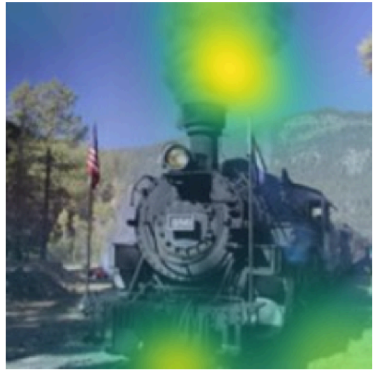
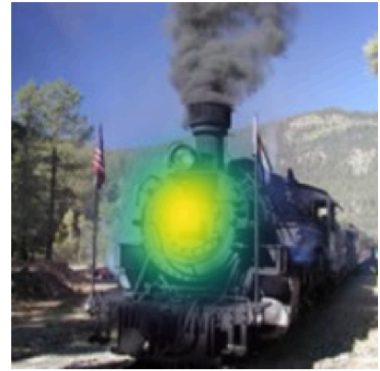
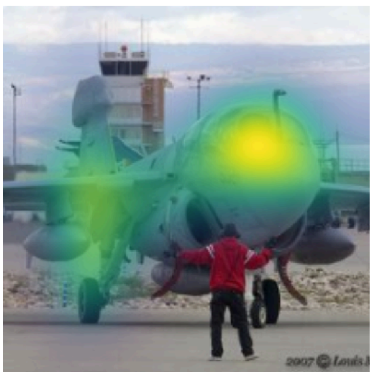
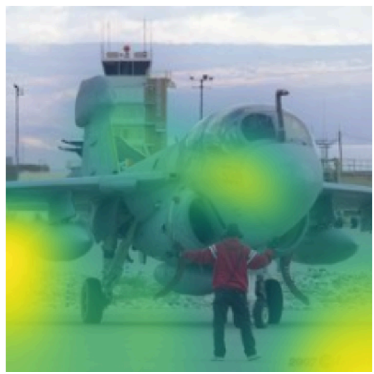
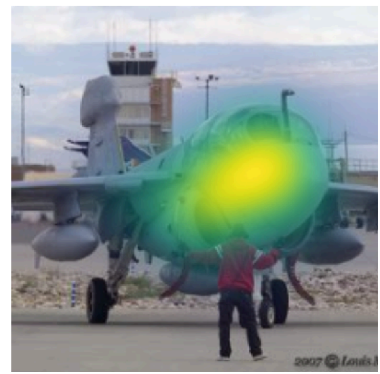

**Figure 3. DNNs can be trained to adopt human-like visual strategies.** (Left) ClickMe [88] maps identify pixels humans believe are important for recognizing these object images. (Middle and Right) Decision attribution maps for the same images from a ViT [115] and Harmonized ViT [89] highlight pixels each model uses for recognition.

Without harmonization, ViTs often rely on contextual features like watermarks and backgrounds that are less important to humans.

Vision science has always developed models at smaller scales than the frontier of artificial intelligence. This is partially because of the academic roots of vision science, partially because of a well-founded desire to lean on reductionism to truly understand how vision works, and partially because attempts at incorporating biological inspiration into DNNs have been hamstrung by implementations that are poorly suited for GPUs. For example, most attempts at biologically-inspired DNNs have focused on inducing architectural constraints like recurrence [34,61,116–118] and different forms of feedback [59,60] that are not explicitly included in DNNs but known to play key roles in primate vision [119–123]. While we believe these approaches are important for Neuroscience and especially for constraining model hypothesis spaces in small data settings, the methods used for implementation are undeniably challenging to scale [118,124] and it is possible that the induced computational strategies could be learned by a less-constrained DNN trained with the "right" data and objective [125]. Thus, it may be that a more effective approach to reverse-engineering vision than hand-designing small-scale recurrent DNNs could be to train DNNs at large scales with approximations of the kinds of data and routines that shape biological visual systems. With that said, it is challenging to imagine how certain types of biological features, like topography [20–22] or neuronal noise [126], could emerge from a DNN without being induced.

We propose that better models of primate vision will come from training DNNs with biologically-grounded data diets and learning principles (see *Outstanding Questions*). While this proposal has been made before [127], the unprecedented computational resources available today now enable this approach to be scaled far beyond previous attempts—a scale that we believe is necessary for breaking trends and building better DNN models of biological vision. Today's GPU systems can support systematic exploration of biologically-aligned optimization objectives while enabling models to interact with naturalistic, dynamic environments rather than passive learning from static images [128,129]. Modern computer vision techniques have essentially eliminated the gap between simulated and real images [130,131], and are now extending into video generation [132]. If spaces of objective functions can be parameterized, then the development of DNNs that properly simulate biological vision is now primarily constrained by resources rather than ideas. We believe this approach could significantly impact vision science, yielding computational models that not only achieve human-level performance but also serve as scientific instruments for elucidating biological vision mechanisms.

**Outstanding questions**

DNNs can be forced into alignment with primate vision by training them with human psychophysics data via Harmonization. Is it possible to achieve similar—or better—outcomes by training DNNs with data diets and learning principles that are more similar to those that govern the development of biological vision? We outline several critical questions:

1. **How can we create more biologically realistic training regimens for DNNs?** Current DNNs typically train on massive, static image collections that bear little resemblance to the temporally structured, multimodal experiences shaping biological vision during development. How can we bridge this gap? One new direction is training DNNs *within* more ecologically valid, dynamic environments [133]. For instance, generative models like Neural Radiance Fields [130] or Gaussian Splatting [131] can create synthetic developmental "diets:" controllable, spatiotemporally coherent visual experiences that mimic aspects of how infants might learn about the world (e.g., observing objects from different viewpoints over time). Earlier and smaller-scale efforts along these lines have shown mixed results [42,134–138]. However, we suspect that data generation frameworks that can produce data at the scale needed to train today's highest-performing DNNs may lead to different outcomes.
2. **What learning principles will align DNNs with biological vision?** It is well-known that standard approaches to train DNNs for classification do not resemble how humans learn about the world [139]. Moreover, while there have been many different self-supervised learning rules released over recent years, none of these have led to better models for primate vision (Fig. 1). The objective functions used in today's DNNs are designed with static image data in mind. What happens when we move our models to dynamic and embodied data collection? What types of objectives and data curricula might cause DNNs to learn more human-like visual representations with these types of data? We suspect that the answers to these questions might not only mold DNNs into better models of human vision, but the principles that achieve these outcomes may also advance our understanding of biological intelligence.
3. **What role will biological constraints play in the search for data diets and learning principles?** It is challenging to imagine how certain features of brains—like topographical organization and neural noise—could emerge without being explicitly induced. Should these features be integrated into our search for optimal data diets and learning principles, or can they be incorporated after developing core visual processing systems? We expect that resolving this question depends on the functional role of these features, an understanding that is rapidly evolving in neuroscience. Determining which constraints are fundamental principles versus byproducts of biological implementation will be crucial for developing truly brain-like artificial vision systems.

**Glossary**

**Attribution maps**: Visualizations that highlight regions of an input image that influence a neural network's predictions, providing insights into model decision-making processes and feature importance for specific classifications.

> **Convolutional Neural Networks:** Deep neural network models that measure the similarity between a set of weights and image content, locally, at every position within an image.
>
> **Deep Neural Network (DNN)**: A computational model consisting of multiple processing layers that learn representations of data with multiple levels of abstraction, inspired by the neural networks in the brain.
>
> **ImageNet**: A large-scale dataset of labeled images spanning thousands of object categories, widely used as a benchmark for training and evaluating object recognition systems.
>
> **Inferotemporal Cortex (IT/ITC)**: A region in the primate visual cortex responsible for high-level visual processing, particularly object recognition and categorization.
>
> **Neural alignment**: The degree to which neural activity patterns in trained models correlate with or predict neural responses in biological systems when presented with the same stimuli.
>
> **Self-supervised training:** Training models with unlabeled data by creating supervisory signals from the data itself rather than annotations. In computer vision, this typically involves solving pretextual tasks, such as filling in masked image regions.
>
> **Transformer**: A deep learning architecture that relies on self-attention mechanisms to process and relate different positions in input sequences, initially developed for natural language processing but later adapted for computer vision tasks.
>
> **Vision Transformer (ViT)**: A type of transformer architecture adapted for computer vision tasks that processes images as sequences of patches rather than through traditional convolutional operations.
>
> **V4**: An intermediate visual area in the ventral stream of the primate visual cortex that is thought to process complex visual features like shape and color, and which serves as a bridge between early visual areas and higher-level object recognition regions.

## Acknowledgements

We are grateful to Dan Yamins, Martin Schrimpf, and Lakshmi Govindarajan for their help with editing and refining this paper. This work is supported by NSF (IIS-2402875), ONR (N00014-24-1-2026), and the ANR-3IA Artificial and Natural Intelligence Toulouse Institute (ANR-19-PI3A-0004).

## References

1. Yamins, D.L.K. *et al.* (2014) Performance-optimized hierarchical models predict neural responses in higher visual cortex. *Proc. Natl. Acad. Sci. U. S. A.* 111, 8619–8624
2. Yamins, D.L. *et al.* (2013) Hierarchical Modular Optimization of Convolutional Networks Achieves Representations Similar to Macaque IT and Human Ventral Stream. *Adv. Neural Inf. Process. Syst. NeurIPS*
3. DiCarlo, J.J. *et al.* (2012) How does the brain solve visual object recognition? *Neuron* 73, 415–434

4. Hilton, J. *et al.* (2023) Scaling laws for single-agent reinforcement learning *arXiv [cs.LG]*
5. Tay, Y. *et al.* (2022) Scaling Laws vs Model Architectures: How does Inductive Bias Influence Scaling? *arXiv [cs.LG]*
6. Kaplan, J. *et al.* (2020) Scaling Laws for Neural Language Models *arXiv [cs.LG]*
7. Zhai, X. *et al.* (2021) Scaling Vision Transformers *arXiv [cs.CV]*
8. Vaswani, A. *et al.* (2017) Attention is all you need. *Adv. Neural Inf. Process. Syst. NeurIPS*
9. Alabdulmohsin, I. *et al.* (2023) Getting ViT in Shape: Scaling Laws for Compute-Optimal Model Design *arXiv [cs.CV]*
10. Henighan, T. *et al.* (2020) Scaling Laws for Autoregressive Generative Modeling *arXiv [cs.LG]*
11. Linsley, D. *et al.* (2024) The 3D-PC: a benchmark for visual perspective taking in humans and machines. *Int. Conf. Learn. Represent. ICLR*
13. McCoy, R.T. *et al.* (2023) Embers of autoregression: Understanding large language models through the problem they are trained to solve *arXiv [cs.CL]*
14. Zuo, Y. *et al.* (2024) Towards foundation models for 3D vision: How close are we? *arXiv [cs.CV]*
15. Wu, Z. *et al.* (2023) Reasoning or reciting? Exploring the capabilities and limitations of language models through counterfactual tasks *arXiv [cs.CL]*
16. Govindarajan, L. *et al.* (2024) Flexible context-driven sensory processing in dynamical vision models. *Adv. Neural Inf. Process. Syst. NeurIPS*
17. Richards, B.A. *et al.* (2019) A deep learning framework for neuroscience. *Nat. Neurosci.* 22, 1761–1770
18. Kell, A.J.E. *et al.* (2018) A Task-Optimized Neural Network Replicates Human Auditory Behavior, Predicts Brain Responses, and Reveals a Cortical Processing Hierarchy. *Neuron* 98, 630–644.e16
19. Schiatti, L. *et al.* (2023) Modeling Visual Impairments with Artificial Neural Networks: a Review. in *Proceedings of the IEEE/CVF International Conference on Computer Vision*, pp. 1987–1999
20. Rathi, N. *et al.* (2024) TopoLM: brain-like spatio-functional organization in a topographic language model *arXiv [cs.CL]*
21. Margalit, E. *et al.* (2024) A unifying framework for functional organization in early and higher ventral visual cortex. *Neuron* 112, 2435–2451.e7
22. Schrimpf, M. *et al.* (2024) Do topographic deep ANN models of the primate ventral stream predict the perceptual effects of direct IT cortical interventions? *bioRxiv*, 2024.01.09.572970
23. Guo, C. *et al.* (2022) Adversarially trained neural representations are already as robust as biological neural representations. *Int. Conf. Mach. Learn. ICML*
24. Bashivan, P. *et al.* (2019) Neural population control via deep image synthesis. *Science* 364, eaav9436
25. Kar, K. *et al.* (2019) Evidence that recurrent circuits are critical to the ventral stream's execution of core object recognition behavior. *Nat. Neurosci.* 22, 974–983
26. Cadena, S.A. *et al.* (2024) Diverse task-driven modeling of macaque V4 reveals functional specialization towards semantic tasks. *PLoS Comput. Biol.* 20, e1012056
27. Willeke, K.F. *et al.* (2022) The Sensorium competition on predicting large-scale mouse primary visual cortex activity *arXiv [q-bio.NC]*
28. Konkle, T. and Alvarez, G.A. (2022) A self-supervised domain-general learning framework for human ventral stream representation. *Nat. Commun.* 13, 491
29. Conwell, C. *et al.* (2024) A large-scale examination of inductive biases shaping high-level visual representation in brains and machines. *Nat. Commun.* 15, 9383
30. Willeke, K.F. *et al.* (2023) Deep learning-driven characterization of single cell tuning in primate visual area V4 unveils topological organization *bioRxiv*, 2023.05.12.540591

31. Cadena, S.A. *et al.* (2019) How well do deep neural networks trained on object recognition characterize the mouse visual system? In *Real Neurons & Hidden Units: Future directions at the intersection of neuroscience and artificial intelligence @ NeurIPS 2019*
32. Safarani, S. *et al.* (2021) Towards robust vision by multi-task learning on monkey visual cortex. *Adv. Neural Inf. Process. Syst. NeurIPS*
33. Linsley, D. *et al.* (2023) Performance-optimized deep neural networks are evolving into worse models of inferotemporal visual cortex. *Adv. Neural Inf. Process. Syst. NeurIPS*
34. Kietzmann, T.C. *et al.* (2019) Recurrence is required to capture the representational dynamics of the human visual system. *Proc. Natl. Acad. Sci. U. S. A.* 116, 21854–21863
35. Mehrer, J. *et al.* (2021) An ecologically motivated image dataset for deep learning yields better models of human vision. *Proc. Natl. Acad. Sci. U. S. A.* 118
36. Khaligh-Razavi, S.-M. and Kriegeskorte, N. (2014) Deep supervised, but not unsupervised, models may explain IT cortical representation. *PLoS Comput. Biol.* 10, e1003915
37. Cadieu, C.F. *et al.* (2014) Deep neural networks rival the representation of primate IT cortex for core visual object recognition. *PLoS Comput. Biol.* 10, e1003963
38. Hong, H. *et al.* (2016) Explicit information for category-orthogonal object properties increases along the ventral stream. *Nat. Neurosci.* 19, 613–622
39. Cichy, R.M. *et al.* (2016) Comparison of deep neural networks to spatio-temporal cortical dynamics of human visual object recognition reveals hierarchical correspondence. *Sci. Rep.* 6, 27755
40. Dobs, K. *et al.* (2023) Behavioral signatures of face perception emerge in deep neural networks optimized for face recognition. *Proc. Natl. Acad. Sci. U. S. A.* 120, e2220642120
41. Dobs, K. *et al.* (2022) Brain-like functional specialization emerges spontaneously in deep neural networks. *Sci. Adv.* 8, eabl8913
42. Zhuang, C. *et al.* (2021) Unsupervised neural network models of the ventral visual stream. *Proc. Natl. Acad. Sci. U. S. A.* 118
43. Mineault, P. *et al.* (2021) Your head is there to move you around: Goal-driven models of the primate dorsal pathway. *Adv. Neural Inf. Process. Syst. NeurIPS*
44. Cueva, C.J. and Wei, X.-X. (2018) Emergence of grid-like representations by training recurrent neural networks to perform spatial localization. *Int. Conf. Learn. Represent. ICLR*
45. Rajalingham, R. *et al.* (2015) Comparison of Object Recognition Behavior in Human and Monkey. *J. Neurosci.* 35, 12127–12136
46. Rajalingham, R. *et al.* (2018) Large-scale, high-resolution comparison of the core visual object recognition behavior of humans, monkeys, and state-of-the-art deep artificial neural networks. *J. Neurosci.* 38, 7255–7269
47. Dehghani, M. *et al.* (2023) Scaling Vision Transformers to 22 Billion Parameters*arXiv [cs.CV]*
48. Geirhos, R. *et al.* (2021) Partial success in closing the gap between human and machine vision*arXiv [cs.CV]*
49. Geirhos, R. *et al.* (2018) Generalisation in humans and deep neural networks. *Adv. Neural Inf. Process. Syst. NeurIPS*
50. Geirhos, R. *et al.* (2020) Beyond accuracy: quantifying trial-by-trial behaviour of CNNs and humans by measuring error consistency. *Adv. Neural Inf. Process. Syst. NeurIPS*
51. Hermann, K. *et al.* (2020) The Origins and Prevalence of Texture Bias in Convolutional Neural Networks. *Adv. Neural Inf. Process. Syst. NeurIPS*
52. Battleday, R.M. *et al.* (2021) From convolutional neural networks to models of higher-level cognition (and back again). *Ann. N. Y. Acad. Sci.* 1505, 55–78
53. Mahner, F.P. *et al.* (2024) Dimensions underlying the representational alignment of deep neural networks with humans *arXiv [cs.CV]*
54. Kümmerer, M. *et al.* (2022) DeepGaze III: Modeling free-viewing human scanpaths with deep learning. *J. Vis.* 22, 7

55. Kümmerer, M. *et al.* (2016) DeepGaze II: Reading fixations from deep features trained on object recognition *arXiv [cs.CV]*
56. Danier, D. *et al.* (2024) DepthCues: Evaluating monocular depth perception in large vision models *arXiv [cs.CV]*
57. Bonnen, T. *et al.* (2024) Evaluating multiview object consistency in humans and image models. *Neural Inf Process Syst* abs/2409.05862, 43533–43548
58. Malhotra, G. *et al.* (2022) Feature blindness: A challenge for understanding and modelling visual object recognition. *PLoS Comput. Biol.* 18, e1009572
59. Linsley, D. *et al.* (2019) Recurrent neural circuits for contour detection. *Int. Conf. Learn. Represent. ICLR*
60. Kim*, J. *et al.* (2020) Disentangling neural mechanisms for perceptual grouping. *Int. Conf. Learn. Represent. ICLR*
61. Linsley, D. *et al.* (2018) Learning long-range spatial dependencies with horizontal gated recurrent units. *Adv. Neural Inf. Process. Syst. NeurIPS*
62. Linsley, D. *et al.* (2021) Tracking Without Re-recognition in Humans and Machines. *Adv. Neural Inf. Process. Syst. NeurIPS*
63. Muzellec, S. *et al.* (2024) Tracking objects that change in appearance with phase synchrony. *Int. Conf. Learn. Represent. ICLR*
64. Doerig, A. *et al.* (2020) Crowding reveals fundamental differences in local vs. global processing in humans and machines. *Vision Res.* 167, 39–45
65. Bowers, J.S. *et al.* (2022) Deep Problems with Neural Network Models of Human Vision. *Behav. Brain Sci.*
66. Feather, J. *et al.* (2025) Brain-model evaluations need the NeuroAI Turing Test *arXiv [q-bio.NC]*
67. Feather, J. *et al.* (2019) Metamers of neural networks reveal divergence from human perceptual systems. *Adv. Neural Inf. Process. Syst. NeurIPS*
68. Feather, J. *et al.* (2023) Model metamers reveal divergent invariances between biological and artificial neural networks. *Nat. Neurosci.* 26, 2017–2034
69. McIntosh, L.T. *et al.* (2016) Deep Learning Models of the Retinal Response to Natural Scenes. *Adv. Neural Inf. Process. Syst. NeurIPS*
70. Tanaka, H. *et al.* (2019) Revealing computational mechanisms of retinal prediction via model reduction. In *Real Neurons & Hidden Units: Future directions at the intersection of neuroscience and artificial intelligence @ NeurIPS 2019*
71. Tanaka, H. *et al.* (2019) From deep learning to mechanistic understanding in neuroscience: the structure of retinal prediction. *Adv. Neural Inf. Process. Syst. NeurIPS*
72. Maheswaranathan, N. *et al.* (2023) Interpreting the retinal neural code for natural scenes: From computations to neurons. *Neuron* 111, 2742–2755.e4
73. Sinz, F. *et al.* (2018) Stimulus domain transfer in recurrent models for large scale cortical population prediction on video. *Adv. Neural Inf. Process. Syst. NeurIPS*
74. Walker, E.Y. *et al.* (2019) Inception loops discover what excites neurons most using deep predictive models. *Nat. Neurosci.* 22, 2060–2065
75. Fu, J. *et al.* (2024) Pattern completion and disruption characterize contextual modulation in the visual cortex *bioRxiv*, 2023.03.13.532473
76. Lurz, K.-K. *et al.* (2020) Generalization in data-driven models of primary visual cortex*bioRxiv*, *bioRxiv*, 2020.10.05.326256
77. Cichy, R.M. *et al.* (2019) The algonauts project. *Nat. Mach. Intell.* 1, 613–613
78. Hebart, M.N. *et al.* (2023) THINGS-data, a multimodal collection of large-scale datasets for investigating object representations in human brain and behavior. *Elife* 12
79. Schrimpf, M. *et al.* (2020) Integrative Benchmarking to Advance Neurally Mechanistic Models of Human Intelligence. *Neuron* 108, 413–423
80. Schrimpf, M. *et al.* (2018) Brain-Score: Which Artificial Neural Network for Object

Recognition is most Brain-Like? *bioRxiv*
81. Gokce, A. and Schrimpf, M. (2024) Scaling laws for task-optimized models of the primate visual ventral stream *arXiv [cs.LG]*
82. Majaj, N.J. *et al.* (2015) Simple Learned Weighted Sums of Inferior Temporal Neuronal Firing Rates Accurately Predict Human Core Object Recognition Performance. *J. Neurosci.* 35, 13402–13418
83. Deng, J. *et al.* (2009) ImageNet: A large-scale hierarchical image database. in *2009 IEEE Conference on Computer Vision and Pattern Recognition*, pp. 248–255
84. Azeglio, S. *et al.* (2022) Improving Neural Predictivity in the Visual Cortex with Gated Recurrent Connections. In *Brain-Score Workshop*
85. Arcaro, M.J. *et al.* (2020) The neurons that mistook a hat for a face. *Elife* 9
86. Ridnik, T. *et al.* (2020) TResNet: High Performance GPU-Dedicated Architecture *arXiv [cs.CV]*
87. Shankar, V. *et al.* (2020) Evaluating Machine Accuracy on ImageNet. *Int. Conf. Mach. Learn. ICML*
88. Linsley, D. *et al.* (2019) Learning what and where to attend. *Int. Conf. Learn. Represent. ICLR*
89. Fel*, T. *et al.* (2022) Harmonizing the object recognition strategies of deep neural networks with humans. *Adv. Neural Inf. Process. Syst. NeurIPS*
90. Geirhos, R. *et al.* (2020) Shortcut learning in deep neural networks. *Nature Machine Intelligence* 2, 665–673
91. Shahamatdar, S. *et al.* (2024) Deceptive learning in histopathology. *Histopathology* DOI: 10.1111/his.15180
92. Xie, C. *et al.* (2018) Feature denoising for improving adversarial robustness *arXiv [cs.CV]*
93. timm. Available: https://huggingface.co/docs/timm/en/index.
94. Ridnik, T. *et al.* (2021) ImageNet-21K Pretraining for the Masses *arXiv [cs.CV]*
95. Oquab, M. *et al.* (2023) Dinov2: Learning robust visual features without supervision. *arXiv [cs.CV]*
96. Radford, A. *et al.* (2021) Learning Transferable Visual Models From Natural Language Supervision. *Int. Conf. Mach. Learn. ICML*
97. He, K. *et al.* (2015) Deep Residual Learning for Image Recognition *arXiv [cs.CV]*
98. He, K. *et al.* (2016) Identity Mappings in Deep Residual Networks *arXiv [cs.CV]*
99. Fang, Y. *et al.* (2023) EVA-02: A Visual Representation for Neon Genesis *arXiv [cs.CV]*
100. Howard, A. *et al.* (2019) Searching for MobileNetV3 *arXiv [cs.CV]*
101. Hendrycks, D. and Dietterich, T. (2019) Benchmarking neural network robustness to common corruptions and perturbations *arXiv [cs.LG]*
102. Wang, H. *et al.* (2019) Learning robust global representations by penalizing local predictive power *arXiv [cs.CV]*
103. Barbu, A. *et al.* (2019) ObjectNet: A large-scale bias-controlled dataset for pushing the limits of object recognition models. *Adv. Neural Inf. Process. Syst. NeurIPS*
104. Schuhmann, C. *et al.* (2022) LAION-5B: An open large-scale dataset for training next generation image-text models. *Adv. Neural Inf. Process. Syst. NeurIPS Datasets and Benchmarks Track*
105. Ullman, S. *et al.* (2016) Atoms of recognition in human and computer vision. *Proc. Natl. Acad. Sci. U. S. A.* 113, 2744–2749
106. Simonyan, K. and Zisserman, A. (2014) Very Deep Convolutional Networks for Large-Scale Image Recognition *arXiv [cs.CV]*
107. Funke, C.M. *et al.* (2021) Five points to check when comparing visual perception in humans and machines. *J. Vis.* 21, 16
108. Linsley, D. *et al.* (2017) What are the Visual Features Underlying Human Versus Machine Vision? *2017 IEEE International Conference on Computer Vision Workshops (ICCVW)*, pp.

2706–2714
109. Fel, T. *et al.* (2022) CRAFT: Concept Recursive Activation FacTorization for Explainability *arXiv [cs.CV]*
110. Fel, T. *et al.* (2021) Look at the Variance! Efficient Black-box Explanations with Sobol-based Sensitivity Analysis. *Adv. Neural Inf. Process. Syst. NeurIPS*
111. Fel, T. *et al.* (2023) Unlocking Feature Visualization for Deeper Networks with MAgnitude Constrained Optimization. *Adv. Neural Inf. Process. Syst. NeurIPS*
112. Simonyan, K. *et al.* (2013) Deep Inside Convolutional Networks: Visualising Image Classification Models and Saliency Maps *arXiv [cs.CV]*
113. Linsley, D. *et al.* (2023) Adversarial alignment: Breaking the trade-off between the strength of an attack and its relevance to human perception *arXiv [cs.CV]*
114. Fel*, T. *et al.* (2022) Harmonizing the object recognition strategies of deep neural networks with humans. *Adv. Neural Inf. Process. Syst. NeurIPS*
115. Dosovitskiy, A. *et al.* (2020) An Image is Worth 16x16 Words: Transformers for Image Recognition at Scale *arXiv [cs.CV]*
116. Nayebi, A. *et al.* (2018) Task-Driven Convolutional Recurrent Models of the Visual System. *Adv. Neural Inf. Process. Syst. NeurIPS*
117. Spoerer, C.J. *et al.* (2017) Recurrent Convolutional Neural Networks: A Better Model of Biological Object Recognition. *Front. Psychol.* 8, 1551
118. Linsley, D. *et al.* (2020) Stable and expressive recurrent vision models. *Adv. Neural Inf. Process. Syst. NeurIPS*
119. Gilbert, C.D. and Li, W. (2013) Top-down influences on visual processing. *Nat. Rev. Neurosci.* 14, 350–363
120. Roelfsema, P.R. *et al.* (2000) The implementation of visual routines. *Vision Res.* 40, 1385–1411
121. Roelfsema, P.R. (2005) Elemental operations in vision. *Trends Cogn. Sci.* 9, 226–233
122. Roelfsema, P.R. and Houtkamp, R. (2011) Incremental grouping of image elements in vision. *Atten. Percept. Psychophys.* 73, 2542–2572
123. Jeurissen, D. *et al.* (2016) Serial grouping of 2D-image regions with object-based attention in humans. *Elife* 5, e14320
124. van Bergen, R.S. and Kriegeskorte, N. (2020) Going in circles is the way forward: the role of recurrence in visual inference. *Curr. Opin. Neurobiol.* 65, 176–193
125. Huh, M. *et al.* (2024) The platonic representation hypothesis *arXiv [cs.LG]*
126. Dapello, J. *et al.* (2020) Simulating a Primary Visual Cortex at the Front of CNNs Improves Robustness to Image Perturbations. *Adv. Neural Inf. Process. Syst. NeurIPS*
127. Yamins, D.L.K. and DiCarlo, J.J. (2016) Using goal-driven deep learning models to understand sensory cortex. *Nat. Neurosci.* 19, 356–365
128. Zador, A. *et al.* (2023) Catalyzing next-generation Artificial Intelligence through NeuroAI. *Nat. Commun.* 14, 1597
129. Warren, W.H. (2021) Information Is Where You Find It: Perception as an Ecologically Well-Posed Problem. *Iperception* 12, 20416695211000366
130. Mildenhall, B. *et al.* (2020) NeRF: Representing Scenes as Neural Radiance Fields for View Synthesis *arXiv [cs.CV]*
131. Kerbl, B. *et al.* (2023) 3D Gaussian splatting for real-time radiance Field rendering. *ACM Trans. Graph.* 42, 1–14
132. Wu, G. *et al.* (2023) 4D Gaussian Splatting for real-time dynamic scene rendering *arXiv [cs.CV]*
133. Merel, J. *et al.* (2019) Deep neuroethology of a virtual rodent *arXiv [q-bio.NC]*
134. Vong, W.K. *et al.* (2024) Grounded language acquisition through the eyes and ears of a single child. *Science* 383, 504–511
135. Orhan, E. *et al.* (2020) Self-supervised learning through the eyes of a child. *Adv. Neural Inf.*

*Process. Syst. NeurIPS*

136. Orhan, A.E. and Lake, B.M. (2024) Learning high-level visual representations from a child's perspective without strong inductive biases. *Nat. Mach. Intell.* 6, 271–283
137. Ehsani, K. *et al.* (2018) Who let the dogs out? Modeling dog behavior from visual data *arXiv [cs.CV]*
138. Bar, A. *et al.* (2024) EgoPet: Egomotion and Interaction Data from an Animal's Perspective *arXiv [cs.RO]*
139. Cheng, Y.-A. *et al.* (2024) RTify: Aligning deep neural networks with human behavioral decisions *arXiv [cs.AI]*